\documentclass[preprint,12pt,a4paper]{elsarticle}

\usepackage{amsmath,amssymb}
\usepackage{booktabs}
\usepackage{tabularx}
\usepackage{array}
\usepackage{multirow}
\usepackage{graphicx}
\usepackage{microtype}
\usepackage{threeparttable}
\usepackage{xurl}
\usepackage{setspace}
\usepackage[hidelinks]{hyperref}

\journal{Computer Methods and Programs in Biomedicine}
\newcolumntype{Y}{>{\raggedright\arraybackslash}X}
\begin{document}
\begin{frontmatter}

\title{Cross-dataset transportability of pediatric chest X-ray deep learning across three countries: discrimination, calibration, operating-point failure, and limited-label recovery}

\author[aiub]{Nazim-E-Alam\corref{cor1}}
\cortext[cor1]{Corresponding author.}
\ead{nazimriyadh001@gmail.com}
\address[aiub]{Department of Computer Science, American International University-Bangladesh (AIUB), 408/1 (Old KA 66/1), Kuratoli, Khilkhet, Dhaka 1229, Bangladesh}

\begin{abstract}
\textbf{Background and Objective:} External evaluation of medical-imaging AI is often collapsed into discrimination. We evaluated a computational protocol that separately tests discrimination, probability calibration, fixed operating-point transport, shortcut-associated signal, and limited-label recoverability for pediatric pneumonia classification across datasets from three countries.

\textbf{Methods:} After exact-duplicate removal, 5,824 Guangzhou radiographs supported leakage-controlled source development and internal testing. A frozen three-seed DenseNet121 dual-view ensemble was evaluated zero-shot on BDCXR-3257 from Bangladesh ($n=3,257$) and an untouched harmonized VinDr-PCXR/PediCXR test cohort from Vietnam ($n=1,077$). Matched seed-42 variants tested architectural robustness. Secondary BDCXR analyses used a fixed 651-image adaptation pool and 2,606-image hold-out; 163, 326, and 651 labels represented 5\%, 10\%, and 20\% of complete BDCXR.

\textbf{Results:} Internal AUROC was 0.976 with 95.1\% sensitivity. BDCXR and VinDr-PCXR AUROC were 0.798 and 0.742, while frozen-threshold sensitivity fell to 6.2\% and 0\%. Source-to-BDCXR AUROC degradation occurred for a full-image baseline (0.961 to 0.749), ungated dual-view model (0.977 to 0.766), and gated MixStyle model (0.966 to 0.789). With 163 BDCXR labels, Platt recalibration preserved AUROC while increasing held-out sensitivity to 88.3\%, but specificity was 47.9\% and the alert rate was 78.5\%. Two hundred repeated 163-label fits confirmed sensitivity recovery but substantial specificity variability.

\textbf{Conclusions:} Cross-dataset shifts across countries affected ranking, probability alignment, and source-defined decision behavior differently. Transport studies should evaluate these components separately and quantify the operational burden of apparent recovery.
\end{abstract}

\begin{keyword}
Pediatric chest X-ray \sep Pneumonia \sep Deep learning \sep Domain shift \sep Calibration \sep External testing \sep Model transportability
\end{keyword}

\end{frontmatter}

\section{Introduction}
Deep learning has produced strong within-dataset results for chest-radiograph classification, including pediatric pneumonia, but high benchmark performance does not establish transportability to different populations or acquisition environments \cite{kermany2018,rajpurkar2018,calli2021}. Pediatric radiography is particularly sensitive to distributional change because age-dependent anatomy, positioning, body habitus, disease spectrum, equipment, and interpretation practice can differ across sites. Reviews of pediatric radiology AI continue to identify restricted external testing and uncertain clinical generalizability as major limitations, while recent pediatric pneumonia reviews show continued concentration on the Guangzhou/Kermany benchmark \cite{padash2022,field2023,kamran2026,muringathuparambil2026,rickard2025}. Work in CMPB has likewise emphasized that pediatric pneumonia AI should be assessed beyond a single accuracy estimate and with attention to applicability and reference-standard limitations \cite{dominguez2023}.

External degradation is well documented in radiology. Multi-hospital studies have shown that chest-radiograph models can encode site identity and exhibit large performance changes outside the development environment \cite{zech2018}, and systematic reviews report external decreases for most evaluated radiologic deep-learning systems \cite{yu2022}. Pediatric evidence is more limited but points in the same direction \cite{xin2022}. Such failure cannot be attributed to geography alone: public datasets may differ simultaneously in case mix, age range, acquisition hardware, preprocessing, compression, prevalence, disease definitions, and reference standards.

A second problem is that external performance is often summarized by AUROC. Discrimination, probability calibration, and operating-point behavior answer different questions. AUROC measures ranking across thresholds, calibration measures agreement between probability and observed frequency, and sensitivity/specificity depend on a chosen threshold and the target score distribution \cite{guo2017,vancalster2019}. A monotonic target shift can therefore leave AUROC useful while making a source-derived operating point nearly nonfunctional. Conversely, recalibration can improve probabilities and operating characteristics without improving ranking \cite{hwang2020,kuo2021}. This distinction is important for deployment because an apparently successful recalibration may still generate an unacceptable alert burden.

Imaging models can also exploit unstable acquisition-associated signals. Hospital markers, peripheral regions, text, borders, and other non-anatomic correlates have been implicated as shortcuts in chest-radiograph AI \cite{jabbour2020,degrave2021,ongly2024}. Such stress tests cannot prove causal feature use, but they can establish whether target data contain label-correlated signals outside the intended anatomy.

We therefore evaluated a locked computational transportability protocol that separates five components: discrimination, probability calibration, fixed operating-point behavior, shortcut/acquisition-associated signal, and limited-label recoverability. A Guangzhou model was frozen before complete-cohort testing in Bangladesh and untouched testing in Vietnam. Bangladesh was used for adaptation only after its primary zero-shot result was retained. Architecture-robustness experiments tested whether the transport phenomenon depended on the proposed dual-view implementation. We hypothesized that ranking, probability alignment, and source-defined operating behavior would transport differently, and that limited target labels would repair score alignment before producing large improvements in ranking.

\section{Materials and methods}
\subsection{Study design and cohorts}
\begin{figure*}[t]
    \centering
    \includegraphics[width=0.98\textwidth]{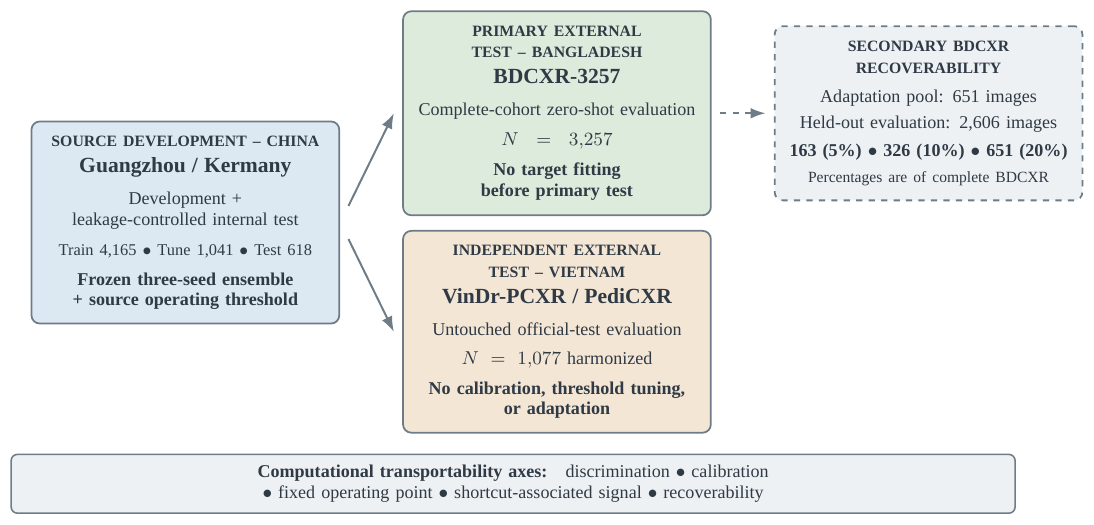}
    \caption{Study design and analysis hierarchy. Guangzhou data were used for model development and internal testing. BDCXR was first evaluated as a complete zero-shot external cohort; only afterward was a secondary target-label recoverability analysis performed on a fixed 651-image adaptation pool and disjoint 2,606-image hold-out. Label budgets of 163, 326, and 651 images correspond to 5\%, 10\%, and 20\% of complete BDCXR. VinDr-PCXR/PediCXR remained untouched and served as a second independent cross-dataset external test.}
    \label{fig:design}
\end{figure*}

This retrospective secondary analysis used de-identified research datasets and did not alter patient care. The analysis hierarchy was fixed to separate zero-shot external testing from later target adaptation. Guangzhou data were used for training, tuning, temperature scaling, threshold selection, and internal testing. BDCXR-3257 was first evaluated in full with the frozen source system. Only after that result was generated and retained was BDCXR partitioned into a fixed 651-image adaptation pool and a disjoint 2,606-image hold-out. The official VinDr-PCXR/PediCXR test split remained untouched: no Vietnam image or label was used for training, calibration, model selection, threshold selection, or adaptation (Fig.~\ref{fig:design}).

The Guangzhou/Kermany source resource contained 5,856 pediatric radiographs labeled normal, bacterial pneumonia, or viral pneumonia \cite{kermany2018}. Bacterial and viral categories were combined for the primary binary radiographic-pneumonia endpoint; etiology labels were used only as an auxiliary source-training signal. BDCXR-3257 contained 3,257 images from Bangladesh (880 normal; 2,377 pneumonia). A verified patient/study identifier was unavailable. VinDr-PCXR/PediCXR provided a separate pediatric DICOM test set from Vietnam \cite{pham2023pedicxr,pham2022vindr}. Before inference, a label-harmonization protocol defined Pneumonia, Brocho-pneumonia, or Pleuro-pneumonia as positive and clean No finding as negative, leaving 1,077 examinations (907 normal; 170 pneumonia-family). Pneumonia-only and isolated pneumonia-family endpoints were also frozen for sensitivity analysis. Table~\ref{tab:datasets} summarizes the analytical roles.

\begin{table*}[t]
\centering
\caption{Cohorts, endpoints, and analytical roles.}
\label{tab:datasets}
\small
\setlength{\tabcolsep}{4pt}
\resizebox{\textwidth}{!}{%
\begin{tabularx}{\textwidth}{@{}p{2.35cm}p{1.55cm}p{2.55cm}p{1.8cm}Y@{}}
\toprule
Cohort & Country & Analytical role & Final $N$ & Key design restriction \\
\midrule
\shortstack[l]{Guangzhou/\\Kermany} & China & Development, tuning, internal test & 5,824 unique & Exact duplicates removed; filename-derived leakage-control groups \\
BDCXR-3257 & Bangladesh & Primary external test; secondary adaptation & 3,257 & No verified patient/study identifier \\
\shortstack[l]{VinDr-PCXR/\\PediCXR} & Vietnam & Untouched external test & 1,077 harmonized & No target fitting; global-diagnosis harmonization frozen before inference \\
\bottomrule
\end{tabularx}
}
\end{table*}

\subsection{Source audit, preprocessing, and model}
Every Guangzhou image was decoded and hashed before model development. Thirty-two exact duplicate images were removed, leaving 5,824 unique radiographs. Filename-derived grouping was performed within disease namespaces because numeric identifiers were reused across bacterial and viral filenames. The final split contained 4,165 training, 1,041 tuning, and 618 internal-test images, with zero exact-hash and derived-group overlap. These groups are leakage-control proxies rather than verified clinical patient identifiers.

Images underwent robust intensity scaling, aspect-ratio-preserving letterboxing, resizing to $320\times320$, three-channel replication, and ImageNet normalization. VinDr DICOM rescale parameters and MONOCHROME1 inversion were handled before the same downstream preprocessing. Lung-conditioned views were generated with the TorchXRayVision anatomical PSPNet and deterministic mask cleanup \cite{cohen2022torchxrayvision}. The classifier used an ImageNet-pretrained DenseNet121 encoder shared across full-radiograph and lung-conditioned views \cite{huang2017densenet}. A learned feature-wise gate combined the two representations, and MixStyle regularization was active only during source training \cite{zhou2021mixstyle}.

For dual-view models, the exact objective that generated the analyzed checkpoints was
\[
L=L_{\mathrm{fused}}+0.25L_{\mathrm{full}}+0.25L_{\mathrm{lung}}+0.10L_{\mathrm{JS}}+0.10L_{\mathrm{etiology}}.
\]
The first three terms were binary cross-entropy losses; $L_{\mathrm{JS}}$ enforced prediction consistency across views, and the etiology cross-entropy was used only for pneumonia images with a valid bacterial/viral source label. No feature-level consistency term was used. Training used AdamW, augmentation, mixed precision, and early stopping on source-tuning performance. Three independent seeds (42, 52, 62) were ensembled by averaging logits. Temperature scaling was fit to ensemble tuning logits only, yielding $T=0.8313$. The primary operating threshold was the highest-specificity source-tuning ROC point attaining at least 90\% sensitivity, yielding 0.9999728. Complete training details are in Supplementary Methods S1--S2.

\subsection{Transport, adaptation, and statistical analyses}
Primary external evaluation reported AUROC and AUPRC, Brier score, NLL, 15-bin ECE, and threshold-dependent sensitivity, specificity, balanced accuracy, predictive values, alert rate, false alerts, and false negatives. AUPRC was interpreted in relation to each cohort's prevalence. To test architecture dependence, matched seed-42 models used the identical source split: conventional full-image DenseNet121, dual view without learned gating or MixStyle, and the gated MixStyle model. Each arm received independent source-only temperature scaling and threshold selection, and external results were not used to choose an architecture.

The secondary BDCXR recoverability analysis used target-label budgets of 163, 326, and 651 images, corresponding to 5\%, 10\%, and 20\% of the complete cohort. Calibration-only methods included threshold adjustment, intercept and temperature scaling, Platt recalibration, and isotonic regression. Representation-update methods included linear probing, head-only, last-block, and full fine-tuning. The prespecified comparison was seed-42 last-block fine-tuning versus matched Platt recalibration at 326 labels. Multi-seed adaptation and five repeated 326-label last-block runs evaluated stability. A separate repeated-recalibration analysis fit Platt models across 200 stratified target-label resamples at 163 and 326 labels; 651-label bootstrap resamples characterized calibrator-fit variability because that budget exhausted the adaptation pool (Supplementary Methods S3 and S6).

Shortcut-associated stress tests re-evaluated BDCXR using full, lung-conditioned, background-only, and border-only views; separate domain classifiers assessed source-versus-target separability. A post-hoc robustness analysis derived a more liberal source-only threshold from Guangzhou tuning. Requests for at least 95\% and 99\% source sensitivity selected the same next ROC point, 0.314311, which was then applied unchanged to all cohorts.

Wilson intervals were used for binomial operating metrics where appropriate. VinDr discrimination and liberal-threshold metrics used 2,000 stratified image-level bootstrap replicates; the paired BDCXR fine-tuning comparison used 2,000 paired bootstrap replicates on the same held-out images. Repeated recalibration reports empirical 2.5th--97.5th percentiles. Because no verified BDCXR patient identifier was available, image-level resampling cannot account for unobserved within-child dependence and may understate uncertainty. Reporting followed CLAIM 2024 and TRIPOD+AI principles \cite{tejani2024claim,collins2024tripod}.

\section{Results}
\subsection{Internal performance and architecture robustness}
After deduplication, the source split contained 1,579 normal and 4,245 pneumonia images. The three-seed ensemble achieved internal-test AUROC 0.9759 and AUPRC 0.9820. At the frozen threshold, sensitivity was 95.1\% (368/387; 95\% CI 92.5--96.8\%), specificity 90.9\% (210/231; 95\% CI 86.5--94.0\%), and balanced accuracy 93.0\%.

The source-to-BDCXR deterioration was not specific to the gated model (Table~\ref{tab:archrobust}). Full-image DenseNet121 decreased from source-test AUROC 0.9606 to 0.7492 on BDCXR; the ungated dual-view model decreased from 0.9772 to 0.7664; and the gated MixStyle model decreased from 0.9655 to 0.7888. Source-defined sensitivity fell from approximately 95\% to 16.2\%, 4.7\%, and 4.4\%, respectively.

\begin{table*}[t]
\centering
\caption{Matched seed-42 architecture robustness from Guangzhou source testing to complete-cohort BDCXR.}
\label{tab:archrobust}
\scriptsize
\setlength{\tabcolsep}{2.5pt}
\resizebox{\textwidth}{!}{%
\begin{tabular}{@{}lrrrrrr@{}}
\toprule
Architecture & Source threshold & Source AUROC & BDCXR AUROC & Source sensitivity & BDCXR sensitivity & BDCXR specificity \\
\midrule
Full-image DenseNet121 & 0.938630 & 0.9606 & 0.7492 & 94.8\% & 16.2\% & 98.1\% \\
Dual view, no gate/MixStyle & 0.999934 & 0.9772 & 0.7664 & 96.1\% & 4.7\% & 100\% \\
Gated dual view + MixStyle & 0.999828 & 0.9655 & 0.7888 & 95.9\% & 4.4\% & 100\% \\
\bottomrule
\end{tabular}
}
\end{table*}

\subsection{Zero-shot external transport}
On complete-cohort BDCXR, AUROC was 0.7984 and AUPRC 0.9116, but only 147/2,377 pneumonia images exceeded the frozen source threshold: sensitivity 6.2\% (95\% CI 5.3--7.2\%), specificity 99.9\% (879/880), and balanced accuracy 53.0\%. The alert rate was 4.5\% and false negatives were 68.5 per 100 radiographs. Calibration also deteriorated (ECE 0.258; Brier 0.268; NLL 1.546).

\begin{figure}[t]
    \centering
    \includegraphics[width=0.85\textwidth]{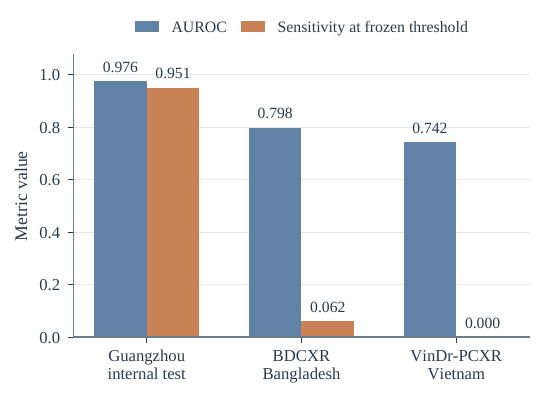}
    \caption{Discrimination versus frozen operating-point transport. AUROC remained above chance at both external sites while sensitivity at the frozen source threshold fell from 95.1\% internally to 6.2\% in BDCXR and 0\% in VinDr-PCXR.}
    \label{fig:transport}
\end{figure}

The untouched VinDr-PCXR cohort independently reproduced the separation between ranking and operating behavior. AUROC was 0.7421 (95\% bootstrap CI 0.7007--0.7815) and AUPRC 0.3960, but none of the 170 pneumonia-family examinations crossed the frozen threshold: sensitivity 0\% (95\% Wilson CI 0--2.2\%), specificity 100\%, and balanced accuracy 50.0\%. ECE was 0.139 and calibration slope 0.196. Pneumonia-only and isolated pneumonia-family definitions yielded AUROC 0.7292 and 0.7339, respectively, with the same 0\% frozen-threshold sensitivity. Figure~\ref{fig:transport} and Table~\ref{tab:transport} summarize the primary transport results.

\begin{table*}[t]
\centering
\caption{Internal and external performance at the frozen source operating point.}
\label{tab:transport}
\small
\resizebox{\textwidth}{!}{%
\begin{tabular}{@{}lrrrrrrrr@{}}
\toprule
Cohort & $N$ & Pneumonia $n$ (\%) & AUROC & AUPRC & Sensitivity & Specificity & Balanced acc. & ECE \\
\midrule
Guangzhou internal test & 618 & 387 (62.6) & 0.976 & 0.982 & 95.1\% & 90.9\% & 93.0\% & 0.152 \\
BDCXR external & 3,257 & 2,377 (73.0) & 0.798 & 0.912 & 6.2\% & 99.9\% & 53.0\% & 0.258 \\
VinDr-PCXR external & 1,077 & 170 (15.8) & 0.742 & 0.396 & 0\% & 100\% & 50.0\% & 0.139 \\
\bottomrule
\end{tabular}
}
\end{table*}

\subsection{Limited-label recovery}
\begin{figure}[t]
    \centering
    \includegraphics[width=0.85\textwidth]{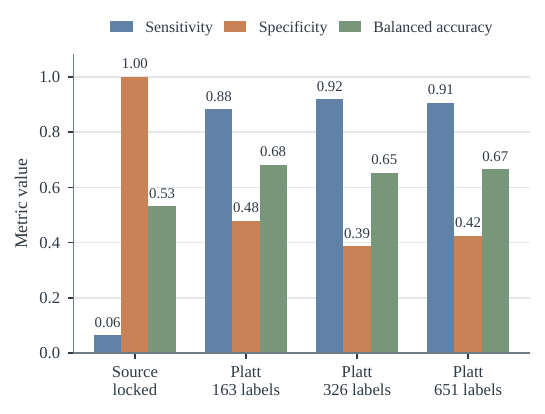}
    \caption{Calibration-only recovery on the locked BDCXR hold-out. Platt recalibration used 163, 326, or 651 labeled adaptation images (5\%, 10\%, or 20\% of complete BDCXR). Because the transform is monotonic, AUROC was preserved while sensitivity, specificity, and calibration changed.}
    \label{fig:recovery}
\end{figure}

On the fixed 2,606-image BDCXR hold-out, source-locked AUROC was 0.7987, sensitivity 6.4\%, specificity 100\%, and ECE 0.256. Platt recalibration was monotonic and therefore preserved ranking. With 163 labels, sensitivity increased to 88.3\% and ECE fell to 0.044, but specificity was 47.9\%, 78.5\% of radiographs generated positive alerts, and there were 14.1 false alerts per 100 examinations. At 326 labels, sensitivity was 91.9\%, specificity 38.6\%, and ECE 0.029; at 651 labels, sensitivity was 90.7\%, specificity 42.5\%, and ECE 0.030 (Table~\ref{tab:recovery}; Fig.~\ref{fig:recovery}).

\begin{figure*}[t]
    \centering
    \includegraphics[width=0.98\textwidth]{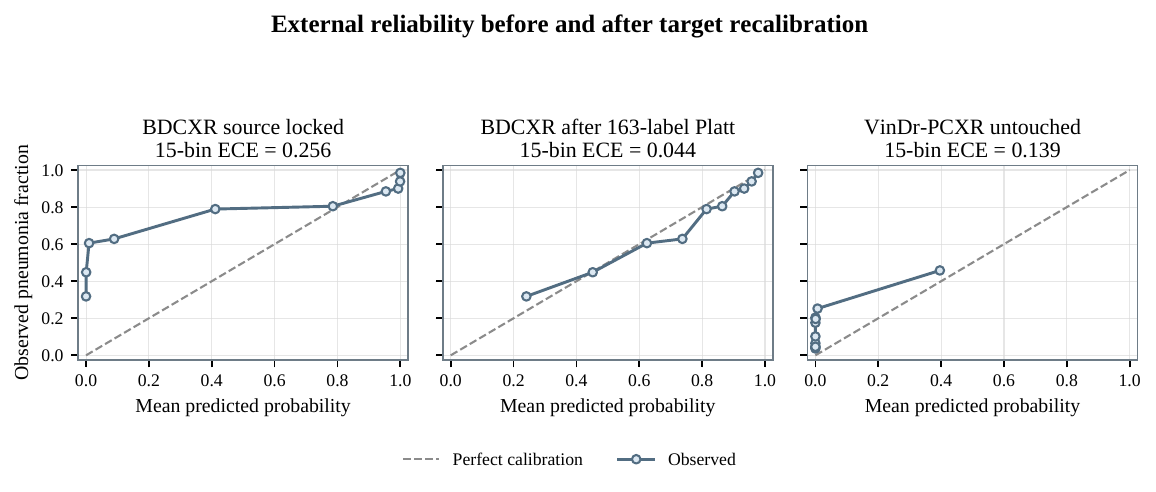}
    \caption{Reliability before and after target recalibration. Displayed points summarize equal-count probability bins for locked BDCXR source predictions, BDCXR after the primary 163-label Platt fit, and untouched VinDr-PCXR predictions. Recalibration substantially improved BDCXR probability alignment but did not establish an acceptable clinical operating policy.}
    \label{fig:reliability}
\end{figure*}

Repeated target-label sampling showed that the direction of calibration recovery was reproducible but the operating point remained sample dependent. Across 200 seed-42 163-label subsamples, mean sensitivity was 91.5\% (empirical 2.5th--97.5th percentile 86.8--95.5\%), mean specificity 37.1\% (21.0--49.3\%), and mean ECE 0.028 (0.017--0.051). At 326 labels, mean sensitivity was 91.5\% (90.0--93.0\%), specificity 38.0\% (32.7--41.1\%), and ECE 0.023 (0.015--0.032). Reliability diagrams show the probability-scale change while also emphasizing that calibration alone does not define an acceptable decision policy (Fig.~\ref{fig:reliability}).

The prespecified 326-label last-block comparison yielded AUROC 0.7907 versus 0.7893 for matched Platt recalibration, a difference of +0.00136 (95\% CI $-0.00117$ to +0.00382; $p=0.293$). Three-seed last-block ensembles reached AUROC 0.8048 with 326 labels and 0.8148 with 651 labels, indicating modest ranking gains as target supervision increased.

\begin{table*}[t]
\centering
\caption{Secondary BDCXR recovery and operational burden on the locked 2,606-image hold-out.}
\label{tab:recovery}
\small
\setlength{\tabcolsep}{4pt}
\resizebox{\textwidth}{!}{%
\begin{tabular}{@{}llrrrrrrrr@{}}
\toprule
Method & Labels & AUROC & Sens. & Spec. & ECE & PPV & NPV & Alert rate & False alerts/100 \\
\midrule
Source locked & 0 & 0.7987 & 0.064 & 1.000 & 0.256 & 1.000 & 0.283 & 0.047 & 0.0 \\
Platt & 163 (5\%) & 0.7987 & 0.883 & 0.479 & 0.044 & 0.821 & 0.603 & 0.785 & 14.1 \\
Platt & 326 (10\%) & 0.7987 & 0.919 & 0.386 & 0.029 & 0.802 & 0.638 & 0.837 & 16.6 \\
Platt & 651 (20\%) & 0.7987 & 0.907 & 0.425 & 0.030 & 0.810 & 0.629 & 0.818 & 15.5 \\
\bottomrule
\end{tabular}
}
\end{table*}

\subsection{Shortcut signal and threshold robustness}
BDCXR pneumonia AUROC remained above chance outside the lungs: 0.7939 on full images, 0.7603 on lung-conditioned images, 0.7185 on background-only content, and 0.6672 on border-only content. Source-versus-BDCXR domain classification was nearly perfect across views, showing that strong acquisition-associated information was present without proving that any specific shortcut caused the observed errors.

\begin{figure}[t]
    \centering
    \includegraphics[width=0.85\textwidth]{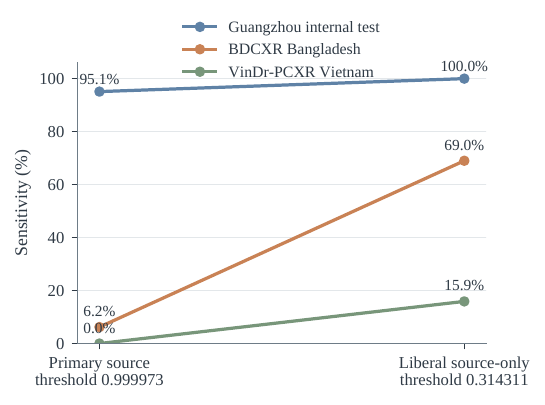}
    \caption{Post-hoc source-threshold robustness. A substantially more liberal operating point derived only from Guangzhou tuning data achieved 100\% internal sensitivity but 69.0\% sensitivity on BDCXR and 15.9\% on VinDr-PCXR.}
    \label{fig:threshold}
\end{figure}

\begin{figure*}[t]
    \centering
    \includegraphics[width=0.90\textwidth]{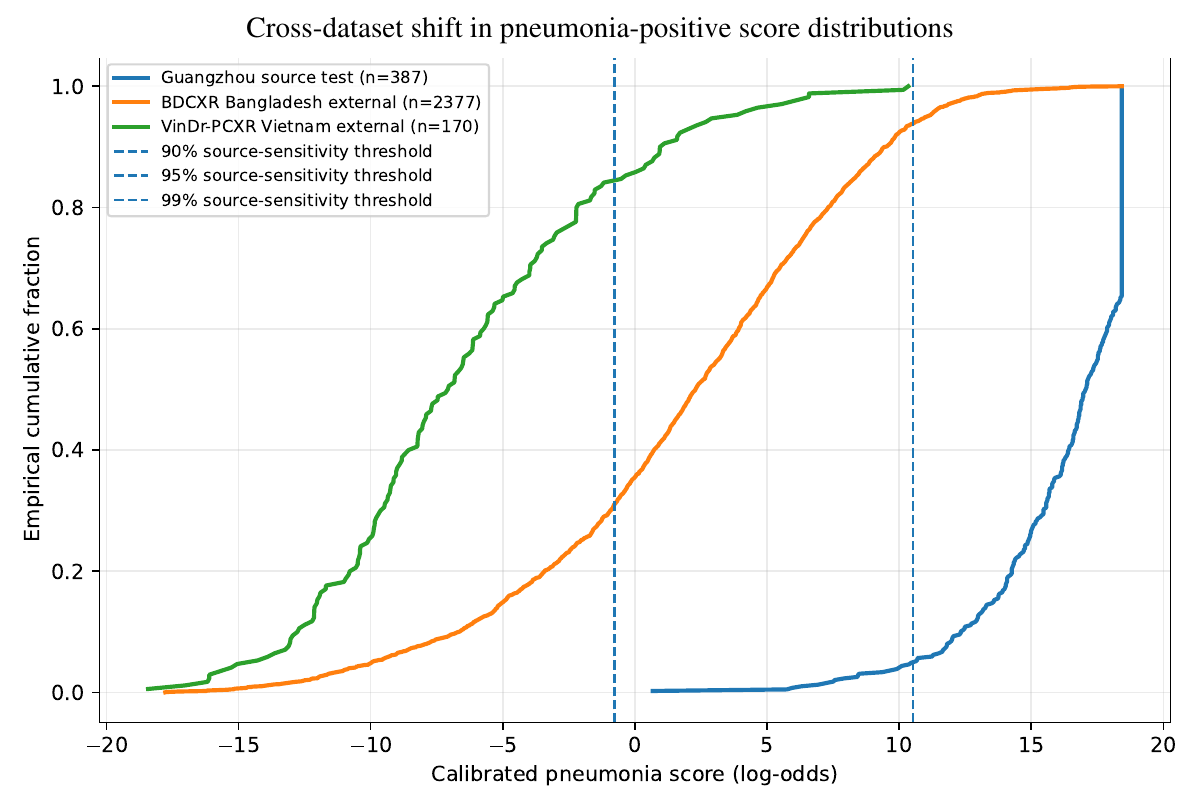}
    \caption{Cross-dataset displacement of calibrated scores among pneumonia-positive examinations. Empirical cumulative distributions are shown for Guangzhou internal-test, BDCXR, and VinDr-PCXR pneumonia-positive examinations on the calibrated log-odds scale. Vertical lines mark source-derived operating thresholds. The figure characterizes score transport and does not attribute the displacement causally to country.}
    \label{fig:scoredist}
\end{figure*}

The liberal source-only threshold increased Guangzhou internal sensitivity to 100\% but produced only 69.0\% sensitivity in BDCXR and 15.9\% in VinDr-PCXR (Table~\ref{tab:threshold}; Fig.~\ref{fig:threshold}). Positive-class score distributions were strongly displaced toward lower values in both target datasets, especially VinDr-PCXR (Fig.~\ref{fig:scoredist}). Thus the unusual primary threshold contributed to the Bangladesh failure but did not explain the larger cross-dataset score shift.

\begin{table*}[t]
\centering
\caption{Sensitivity under the primary and liberal source-defined operating points.}
\label{tab:threshold}
\small
\resizebox{\textwidth}{!}{%
\begin{tabular}{@{}lrrr@{}}
\toprule
Operating point & Guangzhou internal test & BDCXR & VinDr-PCXR \\
\midrule
Primary source threshold (0.9999728) & 95.1\% & 6.2\% & 0\% \\
Liberal source-only threshold (0.314311) & 100\% & 69.0\% & 15.9\% \\
\bottomrule
\end{tabular}
}
\end{table*}

\section{Discussion}
This study demonstrates that cross-dataset transportability of pediatric CXR AI is multidimensional. The principal computational contribution is a locked evaluation protocol that separates discrimination, calibration, fixed operating-point behavior, shortcut/acquisition-associated signal, and limited-label recoverability. Across both external cohorts, the frozen ensemble retained above-chance ranking while source-defined operating behavior failed much more severely. The architecture-robustness analysis showed the same qualitative source-to-BDCXR degradation for conventional full-image DenseNet121 and both dual-view variants, making the transport result less dependent on one implementation. The untouched Vietnam cohort further replicated the central pattern without any target fitting.

The first implication is that AUROC alone can mischaracterize deployment performance. BDCXR AUROC remained 0.798 while frozen-threshold sensitivity was 6.2\%; VinDr-PCXR AUROC remained 0.742 while sensitivity was 0\%. The liberal source-only threshold partly recovered Bangladesh sensitivity but left a large gap in Vietnam, and the positive-score distributions showed pronounced target displacement. The very high primary threshold therefore amplified, but did not create, the operating failure. External evaluation should preserve the distinction between ranking and absolute score behavior rather than treating either AUROC or a single thresholded metric as a complete description \cite{vancalster2019,yu2022}.

The BDCXR intervention ladder provides a second practical distinction: score-scale recovery versus representation recovery. With 163 labels, Platt recalibration preserved AUROC but sharply improved probability alignment and access to a sensitivity-oriented operating region. Repeated label draws confirmed this direction while exposing large specificity variability. Moreover, the resulting alert rate of 78.5\% and false-positive burden show why ``recovery'' should not be equated with clinical readiness. The prespecified fine-tuning comparison produced no convincing AUROC advantage at 326 labels, whereas larger multi-seed adaptation eventually produced only modest ranking gains. When external ranking remains useful, simple recalibration is therefore a reasonable first computational intervention, but it must be evaluated together with workload and error costs \cite{hwang2020,kuo2021}.

The shortcut-associated results provide context rather than causal proof. Background-only and border-only images remained predictive on BDCXR, and source-target domain classification was nearly perfect. These observations are consistent with prior evidence that chest-radiograph AI can exploit hospital and acquisition characteristics \cite{zech2018,jabbour2020,degrave2021,ongly2024}. However, the public datasets do not provide the controlled acquisition metadata or interventions required to identify a specific causal shortcut.

The analysis hierarchy is also important. Once BDCXR was inspected and used for calibration or fine-tuning, those secondary results were no longer untouched external validation. The complete-cohort BDCXR result was therefore retained separately and VinDr-PCXR remained unused for adaptation. This distinction is consistent with contemporary reporting principles that separate development, external testing, and later model updating \cite{tejani2024claim,collins2024tripod}.

The study has several limitations. The three resources differ simultaneously in geography, age distribution, prevalence, equipment, preprocessing/compression, disease spectrum, annotation protocol, and reference standard; the findings therefore demonstrate cross-dataset transport failure across countries, not a causal country effect. Guangzhou leakage groups were reconstructed from filenames and are not verified patient identifiers. Most importantly, BDCXR lacks a verified patient/study identifier, so the adaptation analysis is image-level and its bootstrap intervals cannot account for possible within-child dependence. VinDr-PCXR required endpoint harmonization, although narrower definitions produced the same qualitative operating failure. Finally, the study is retrospective and does not include prospective deployment, local radiologist re-adjudication, or decision-curve analysis. The architecture is intentionally not presented as a novel vision backbone; the contribution is the reproducible computational decomposition of transport failure and its use across multiple datasets and model variants. Prospective multi-institutional studies with verified patient identifiers and prespecified operating costs are needed.

\section{Conclusions}
Across pediatric CXR datasets from China, Bangladesh, and Vietnam, cross-dataset shift affected discrimination, calibration, and source-defined operating behavior differently. Limited target labels consistently improved probability alignment but did not ensure a clinically acceptable operating policy. External medical-imaging AI studies should therefore evaluate discrimination, calibration, operating-point transport, shortcut-associated signal, and recoverability separately, while keeping untouched external testing distinct from later adaptation.

\section*{Ethics statement}
This study was a secondary computational analysis of de-identified research datasets released for research use. No new participants were recruited, no intervention was performed, and no identifiable information was accessed. Ethics approvals, consent procedures, and data-governance conditions for the original collections are described by the respective dataset providers and publications. No additional participant consent was obtained for this secondary analysis.

\section*{CRediT authorship contribution statement}
Nazim-E-Alam: Conceptualization, Methodology, Software, Validation, Formal analysis, Investigation, Data curation, Visualization, Writing -- original draft, Writing -- review \& editing, Project administration.

\section*{Funding}
This research did not receive any specific grant from funding agencies in the public, commercial, or not-for-profit sectors.

\section*{Declaration of competing interest}
The author declares that he has no known competing financial interests or personal relationships that could have appeared to influence the work reported in this paper.

\section*{Data and code availability}
The source datasets are available from their original repositories subject to their licenses and access conditions. A curated reproducibility-code archive is supplied with this submission and contains executable audit, adaptation, external-test, threshold-robustness, and repeated-recalibration workflows together with integrity hashes and derived summaries. Raw radiographs and restricted VinDr-PCXR DICOM data are not redistributed.

\section*{Acknowledgements}
None.

\section*{Declaration of generative AI and AI-assisted technologies in the manuscript preparation process}
During preparation of this work, the author used OpenAI ChatGPT for literature synthesis, code debugging, manuscript structuring, and language refinement. The author reviewed and edited the content, verified numerical claims against preserved analysis outputs, and takes full responsibility for the article. No generative-AI system was used to create or alter scientific image data; figures were produced deterministically from analysis outputs or as vector schematics.

\bibliographystyle{elsarticle-num}
\bibliography{references}

\end{document}